\documentclass[12pt]{iopart}

\usepackage{graphicx}
\usepackage{epsfig}
\usepackage{verbatim}
\usepackage{amssymb,bm}
\usepackage[T1]{fontenc}
\usepackage{eufrak}

\usepackage{bm}

\begin{document}

\newcommand{\be}{\begin{equation}}
\newcommand{\ee}{\end{equation}}
\newcommand{\ba}{\begin{eqnarray}}
\newcommand{\ea}{\end{eqnarray}}
\newcommand{\si}{\sigma_i^z}
\newcommand{\sj}{\sigma_j^z}

\title[Domain walls and chaos in the disordered SOS model] 
{Domain walls and chaos in the disordered SOS model}

\author{K. Schwarz $^1$, A. Karrenbauer $^{2,3}$, G. Schehr $^4$, 
and H. Rieger $^{1,4}$}

\address{
$^1$ Theoretische Physik, Universit\"at des Saarlandes, 66041
  Saarbr\"ucken Germany\\
$^2$ Max Planck Institute for Informatics, Universit\"at des Saarlandes, 
  66041 Saarbr\"ucken Germany\\
$^3$ \'Ecole Polytechnique F\'ed\'erale de Lausanne, Lausanne, Switzerland\\
$^4$ Laboratoire de Physique Th\'eorique, Universit\'e de
  Paris-Sud, 91405 Orsay France 
}   

\begin{abstract}
Domain walls, optimal droplets and disorder chaos at zero temperature
are studied numerically for the solid-on-solid model on a random
substrate. It is shown that the ensemble of random curves represented
by the domain walls obeys Schramm's left passage formula with
$\kappa=4$ whereas their fractal dimension is $d_s=1.25$, and
therefore is {\it not} described by ''Stochastic-Loewner-Evolution''
(SLE). Optimal droplets with a lateral size between $L$ and $2L$ have
the same fractal dimension as domain walls but an energy that
saturates at a value of order ${\cal O}(1)$ for $L \to\infty$ such that
arbitrarily large excitations exist which cost only a small amount of
energy. Finally it is demonstrated that the sensitivity of the ground
state to small changes of order $\delta$ in the disorder is subtle:
beyond a cross-over length scale $L_\delta \sim \delta^{-1}$ the
correlations of the perturbed ground state with the unperturbed ground
state, rescaled by the roughness, are suppressed and approach zero
logarithmically.
\end{abstract}

\maketitle

\section{Introduction}

Domain walls in disordered systems play an important role in
understanding the stability of the ordered phase, the energetics of
large scale excitations, the asymptotic dynamics in and out of
equilibrium as well as the sensitivity to changes of external
parameters. They have been studied quite intensively in the
recent years for Ising spin glasses 
\cite{bray-moore,fisher-huse,rieger-etal,kawashima,hartmann-young,
amoruso-etal,takayama,hartmann}, XY spin glasses \cite{weigel},
random field systems \cite{natter,natter-rev,alava,middleton-fisher}, 
random ferromagnets \cite{natter}, disordered elastic manifolds 
\cite{fisher,blasum,middleton-wall,pfeiffer,middleton-disloc,middleton-drop},
and many others. 

Two domain wall properties are prominent: the first concerns energy
and can be characterized by the scaling behavior of the domain wall
energy with their lateral size, which gives rise to a first, sometimes
universal exponent, the stiffness exponent $\theta$. The second concerns
geometry and gives rise to another, sometimes universal exponent, the
fractal dimension $d_s$, or, in case the domain is not fractal, a roughness
exponent $\zeta$. The interplay between energetics and geometry of the domain
walls (i.e. between stiffness exponent and fractal
dimension) determines how sensitive the system state is to changes of
either external parameters like the temperature or a field, or internal
parameters like small disorder variations. This sensitivity is often
extreme in glassy systems and goes under the name of ``chaos''
\cite{bray-moore}.

Domain walls of glassy systems in two space dimensions represent
fractal curves in the plane and the question arises, whether they fall
into the general classification scheme for ensembles of random curves
described by Stochastic Loewner Evolution (SLE) \cite{bauer,cardy}.
Recently indications were found that domain walls in 2d spin glasses
(at zero temperature) are indeed described by SLE
\cite{amoruso,middleton-doussal}, at least for a Gaussian distribution
of the bonds, but apparently not for binary couplings
\cite{risaus-gusmann}. Also the domain walls in the random-bond Potts
model at the critical point (i.e\ at finite temperature) were found to
be numerically consistent with SLE \cite{potts}. It appears natural to
ask, whether the domain walls in other two-dimensional disordered
systems are potential candidates for a description by SLE.

In this paper we study domain walls and chaos at zero temperature in
the solid-on-solid (SOS) model on a disordered substrate. This is a
numerically convenient representation of a two-dimensional elastic
medium, with scalar displacement field, interacting with quenched
periodic disorder. It has been studied to describe various physical
situations ranging from vortex lattices in
superconductors to incommensurate charge density waves and crystal
growth on a disordered substrate \cite{giam-elastic,nattermann-review,elast-review}. Here we focus on
three questions: 1) are domain walls in this model described by SLE, 2)
what is the relation between size and energy of optimal excitations
(droplets) in this model, 3) does disorder chaos exist in the
ground state of this model ? After a brief summary of what is already
known about the model and a description of the numerical method by
which we compute the ground state and the domain walls these three
issues are studied in separate sections. The paper ends with a
discussion of the results obtained.

\subsection{Model}

We consider the solid-on-solid model on a disordered substrate 
defined by the Hamiltonian 
\be
H=\sum_{(ij)} (h_i-h_j)^2\;,\quad h_i=n_i+d_i\;,
\label{sos}
\ee
with $i \equiv (x_i,y_i) \in {\mathbb Z}^2$. In Eq. (\ref{sos}) the height variables $n_i$ ($i=1,\ldots,N$) take on integer
values $n_i=0,\pm1,\pm2,\ldots$ and the offsets $d_i$ are independent quenched
random variables uniformly distributed between 0 and 1. The sum is
over all nearest neighbor pairs $(ij)$ of a rectangular lattice of
size $L_x\times L_y$ ($L_x=L_y=L$ if not stated otherwise). The
boundary conditions will be specified below in the context of domain
walls. The Hamiltonian in Eq. (\ref{sos}) describes a discrete model of a
two-dimensional elastic medium in a disordered environment.  In the continuum limit, it is described by a sine-Gordon model with
random phase shifts (and in the absence of vortices), the so called
Cardy-Ostlund model \cite{co},  
\be
H_{\rm CO}=\int d^2{\bf r} (\nabla u({\bf r}))^2 
- \lambda\cos(2\pi[u({\bf r})-d({\bf r})]) \;,
\label{cont}
\ee
with a continuous scalar displacement field $u({\bf r}) \in (-\infty,
+\infty)$ and  
quenched random variables $d({\bf r})\in[0,1]$. 
Discretizing the integral and performing the 
infinite strong coupling limit $\lambda\to\infty$ one
recovers (\ref{sos}).

It is well known that this model (\ref{sos}, \ref{cont}) displays a transition between a high temperature phase, $T > T_g = 2/\pi$ 
where the disorder is irrelevant and a low temperature phase below $T_g$, dominated by the disorder. The high-temperature phase
$T>T_g$ is characterized by a logarithmic thermal roughness  
\be
C(r)=\overline{\langle (h_i -h_{i+r})^2\rangle}\sim T \log r\;,
\ee
where $\langle\ldots\rangle$ denotes the thermal average and
$\overline{\ldots}$ the average over the quenched disorder.
The low-temperature or glassy phase is instead ``superrough'', 
characterized by an asymptotically stronger (log-square) 
increase of $C(r)$:
\be
C(r) \sim c(T) \cdot\log^2 r + {\cal O}(\log(r)) \;,
\ee
which means $\zeta = 0$, as expected for a random periodic system. 
Close to $T_g$, a Coulomb Gas renormalization group (RG) analysis to
lowest order gives  
$c(T) \simeq (1-T/T_g)^2/2 \pi^2$ \cite{elast-review}, in rather good agreement with numerical simulations \cite{zeng-tg}. At $T=0$, numerical
simulations give the estimate $c(T=0)  \approx 0.5/(2 \pi)^2 \approx
0.012$ \cite{blasum,zeng-sos}. While earlier studies, based on
``nearly conformal'' field theory \cite{ludwig}, claimed an exact
result for $A(T)$, predicting $A(T=0) = 0$, in clear contradiction
with numerics, a more recent approach based on FRG, incorporating non
analytic operators predicts a non-zero $A(T=0)$ which compares
reasonably with numerics \cite{doussal-schehr}.

For free or periodic boundary conditions, the Hamiltonians (\ref{sos})
and (\ref{cont}) have a discrete 
symmetry, the energy is invariant under a global height (displacement)
shift $n_i\to n_i+\Delta n$ ($u({\bf r})\to u({\bf r})+\Delta n$),
where $\Delta n$ is an arbitrary integer. This symmetry will not be
broken in the low temperature phase of the infinite system and true
long-range order at $T<T_g$ is absent, {\it i.e.} $\overline{\langle
h_i \rangle} = 0$. Concomitantly the model
(\ref{sos}), with free or periodic boundary conditions, has infinitely
many ground states, which differ by a global shift $\Delta
n\in\{\pm1,\pm2,\ldots\}$.

\begin{figure}[t]
\includegraphics[width=\linewidth]{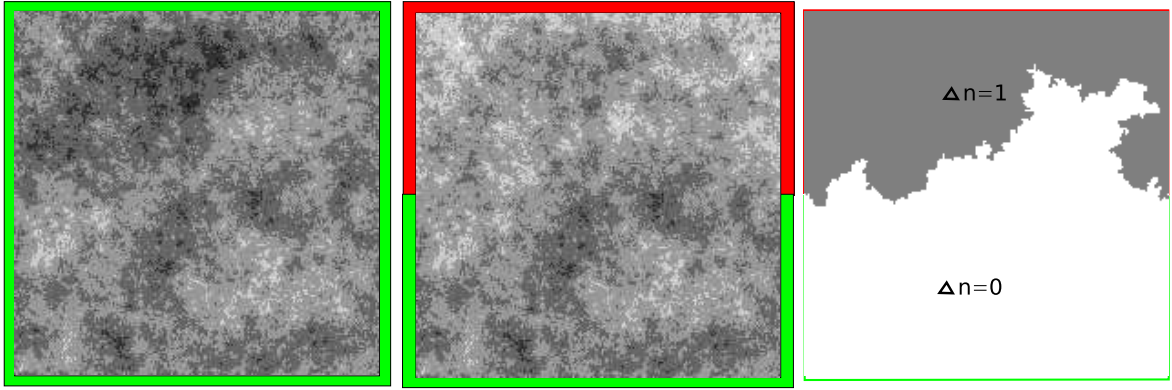} 
\caption{{\bf Left}: Ground state ${\bf n}^0$ of a 
$200\times200$ system with the boundary sites (indicate in green)
fixed to $n_i=0$. The different height values $n_i$ are grey-coded
(dark = low values, bright = high values). {\bf Middle}: Ground state
configuration of the same system as to the left with the upper half of
the boundary sites (indicated in red) fixed to $n_i=1$ and the lower
half (indicated in green) to $n_i=0$. {\bf Right}: Difference plot
between the Left and Middle plots: in the lower white region the
ground state configuration is identical to the corresponding sites in
the left figure, whereas in the upper grey region they differ by
exactly $\Delta n=1$ from the corresponding site in the middle
panel. The border between the white and the grey region is a domain
wall, representing a step in the height profile of the ground state.
}
\label{fig-dw}
\end{figure}

\subsection{Domain walls}

By an appropriate choice of boundary conditions one can force a domain
wall into the system, which is most easily visualized at $T=0$ (c.f.\ Fig.\ \ref{fig-dw}): consider the square
geometry and fix the values of the boundary variables to $n_i = 0$. This yields a unique ground state configuration $n_i^0$. If
one fixes the boundary variables to $n_i=+1$, the corresponding ground
state would be ${n'}_i^0=n_i^0+1$. A domain wall inducing boundary
condition is one, in which the lower half of the boundary values are
fixed to $n_i=0$ and the upper half to $n_i^0=1$. The ground state of
this set-up is then $\tilde{n}_i^0=n_i^0$ in some, mainly the lower
region of the system, and $\tilde{n}_i^0={n'}_i^0$ in the rest - both
region separated by a domain wall of non-trivial shape. 

It turns out that these domain walls are fractal
\cite{blasum,pfeiffer}, which means that their lengths $l_{\rm dw}$
scales with linear system size as 
\be
l_{\rm path}\sim L^{d_s}\;,
\label{dwl}
\ee
with $d_s>d-1=1$. The numerical estimate for $d_s$ is
$d_s=1.27\pm0.02$ \cite{pfeiffer}. Such a fractal scaling of zero-$T$
domain walls is also found for spin glasses, in the 2d EA model with
Gaussian couplings it is $d_{s,SG}=1.27\pm0.01$, and with binary
couplings it is $d_{s,SGB}=1.33\pm0.01$. On the other hand zero-$T$
domain walls in disordered Ising or Potts ferromagnets are rough
(i.e.\ are characterized by algebraic correlations) but not fractal.

The energy for such a domain wall, given by the difference between the
energy of the ground state of the system with the domain wall
inducing boundary conditions and the one with homogeneous boundary
conditions, increases with $L$ logarithmically 
\be
\Delta E\sim \log L\;.
\label{dwe}
\ee
This result, which was obtained by numerical simulations
\cite{blasum}, is consistent with the 
usual scaling relation $\Delta E\sim L^\theta$ together with the exact
result $\theta = d-2+2\zeta = 0$ (thanks to statistical tilt symmetry
\cite{sts}). This 
logarithmic behavior is characteristic of a {\it 
  marginal} glass phase, described by a line 
of fixed point indexed by temperature (which is here marginal in the
RG sense). For comparison the stiffness exponent in 2d (3d) spin 
glasses is $\theta_{SG2d}=-0.28\pm0.01$ and
$\theta_{SG2d}=0.3\pm0.1$ (and thus characterized by a $T=0$ fixed point), whereas for disordered Ising or Potts
ferromagnets $\Delta E \sim L^{d-1}$. 

\subsection{Method}

The ground states of (\ref{sos}), i.e.\ the configuration ${\bf
n}^0=(n_1^0,\ldots,n_N^0)$ with the lowest value for the energy $H[{\bf
n}^0]$ for a given disorder configuration ${\bf d}=(d_1,\ldots,d_N)$,
can be computed very efficiently using a minimum-cost-flow-algorithm
\cite{mincost,blasum,hartmann-rieger}. For the specific details in which 
domain walls are induced in the ground state it is useful to recapitulate the
mapping onto a minimum-cost-flow problem.

After introducing the height-differences $n_{ij}^* = n_i - n_j$
(integer) and $d_{ij}^*=d_j-d_i$ ($\in[-1,+1]$) along the links
$k = (i,j)$ on the dual lattice $G^*$ one obtains a cost (or energy) 
function that lives on the dual lattice
\be
 H[{\bf n}^*] = \sum\limits_k (n_k^* - d_k^*)^2 .
\label{sos2} 
\ee
The configurations ${\bf n}_k^*=(n_1^*,\ldots,n_M^*)$, where $M$ is
the number of links (or bonds) of the original lattice, constitute a
``flow'' on the graph $G^*$. Suppose the original model (\ref{sos}) 
has free boundary conditions. Then the sum of the height differences
along any directed cycle in the original lattice vanishes. Therefore the 
divergence of ${\bf n}^*$ vanishes at all sites~$i$:
\be 
(\nabla \cdot {\bf n}^*)_i=0\;,
\label{con}
\ee
which means that the flow ${\bf n}^*$ on $G^*$, in order to give rise
to a height field ${\bf n}$ on the original lattice $G$ has to be
divergence-less, i.e. without sources or sinks. The problem of
determining the ground state ${\bf n}$ of (\ref{sos}) is thus equivalent to
find the flow ${\bf n}^*$ with the minimum cost (\ref{sos2}) under the
mass-balance constraint (\ref{con}) - i.e. a minimum cost flow
problem, for which there exist very powerful algorithms
\cite{mincost,blasum,hartmann-rieger}. 

\begin{figure}[ht]
\includegraphics[width=\linewidth]{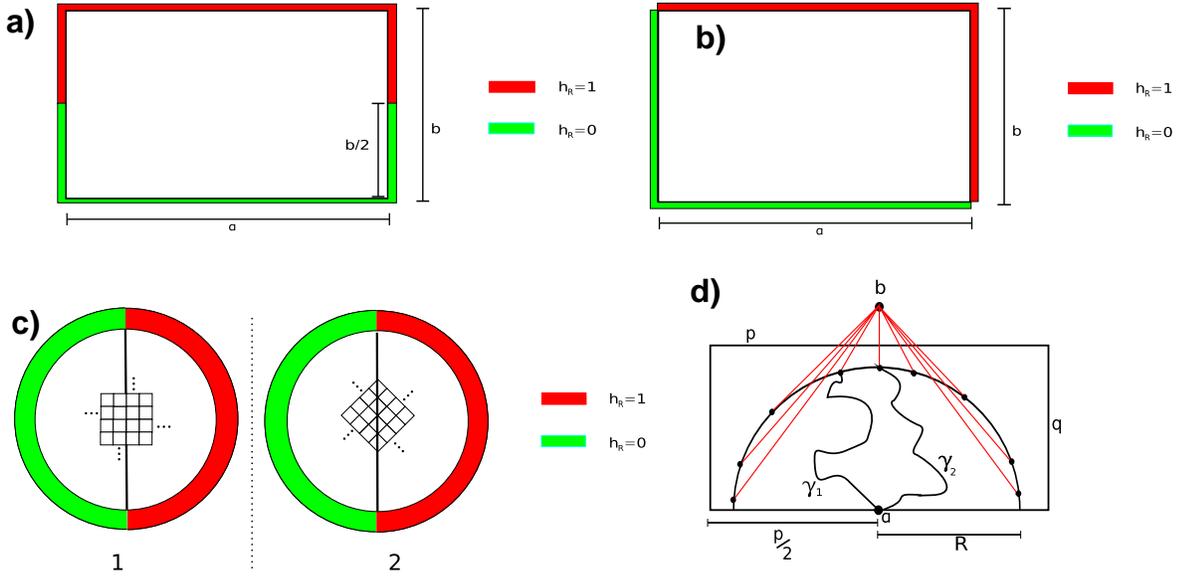} 
\caption{Different geometries and constraints 
on domain walls considered here: a) boundary conditions inducing a
step / domain wall as in Fig.\ \ref{fig-dw}. b) Boundary conditions
inducing a step / domain wall running diagonally from one corner of
the rectangular lattice to the opposite one. c) Boundary conditions
for a circular domain inducing a boundary along the equator
with two different orientations of the underlying lattice. d)
Boundary conditions for a half circle domain and a domain wall 
with one fixed end at the origin and a free end on the outer 
half circle.}
\label{fig-dw2}
\end{figure}

Enforcing one domain wall, or step of height one, into the ground
state of (\ref{sos}) by appropriate boundary condition is then
equivalent to modify the constraint (\ref{con}) at exactly two sites,
the start and end point of the domain wall (see
Fig. \ref{fig-dw2}a-d). As an example consider the
case in which one wants the domain wall to start at the point
$(x,y)=(1,L/2)$ and end at $(x,y)=(L,L/2)$ of a square lattice. Then
one chooses the boundary conditions for $n_i$ as follows (c.f.\ Fig.\
\ref{fig-dw2} a): one fixes the values for $n_i$ at the lower half of the
boundary (i.e.\ at $i=(x,1)$ for $x=1,\ldots,L$ and $i=(1,y)$ and
$(L,y)$ for $y=1,\ldots,L/2$) to $n_i=0$, and the values for $n_i$ at
the upper half of the boundary (i.e.\ at $i=(x,L)$ for $x=1,\ldots,L$
and at $i=(1,y)$ and $(L,y)$ for $y=L/2+1,\ldots,L$) to
$n_i=1$. Translating these boundary conditions for the height
variables ${\bf n}$ into constraints for the flow variables ${\bf
n}^*$ one immediately sees that at the point $(x,y)=(1,L/2)$ and
$(x,y)=(L,L/2)$, where the step in the height profile starts and
terminates, respectively, the constraint (\ref{con}) is modified into
\be 
(\nabla \cdot {\bf n}^*)_{(1,L/2)}=+1\;,\quad
(\nabla \cdot {\bf n}^*)_{(L,L/2)}=-1\;.
\label{con2}
\ee
In other words: the induced step sends a unit of flow from the
starting point of the domain wall, which is the a source of unit
strength, across the sample to the end point, the sink, along an
optimal (minimum cost/energy) path. In what follows we identify 
domain walls immediately with the optimal path for the
extra flow unit defined by the modified mass balance constraints
(\ref{con2}).

With the help of this concept one can then also consider situations in
which the starting point of the domain wall is fixed but the ending
is only forced to be on a specific region of the boundary, opposing
the starting point (see Fig.\ \ref{fig-dw2}d).
Suppose one wants he domain wall to start at
$i_s=(1,L/2)$, and terminate somewhere on the opposing boundary
$i_t=(L,y)$ with $y\in\{1,\ldots,L\}$. Then one introduces an extra
node into the dual graph $G^*$, denoted as the target node, connects
it with bonds of zero cost to all sites on the terminal boundary, and
assigns a sink strength $-1$ to it. The source node is the one closest
to $i_s$ in the dual graph and has source strength $+1$. The minimum
cost flow of this arrangement is then the desired ground state
configuration with a domain wall starting at $i_s$ and ending 
somewhere on the opposite boundary.

\section{Schramm-Loewner evolution (SLE)}

Since the domain walls as defined above represent fractal curves
embedded in a two-dimensional space the question arises whether they
fall into the classification scheme of Schramm-Loewner evolution (SLE)
like loop-erased random walks, percolation hulls, and domain walls at
phase transitions in 2d in the scaling limit
\cite{cardy,bauer}. The necessary (and sufficient) condition 
for a set of random curves connecting two points on the boundary $D$ of
a domain to be described by SLE are 1) the measure for these random
curves has to fulfill a Markov property, 2) the measure has to be
invariant under conformal mappings of the domain. Recently it was
suggested that also domain walls in 2d spin glasses can be described
by SLE \cite{amoruso,middleton-doussal} although neither conformal
invariance is fulfilled for each individual disorder realization nor
the Markov property after disorder averaging. However, conformal
invariance might hold for the disorder averaged model and the Markov
property might be fulfilled almost always in a statistical sense.

A single parameter $\kappa$ parameterizes all SLEs and it is
related to the fractal dimension of the curve via 
\be
d_s=1+\kappa/8\;.
\ee
The parameter $\kappa$ is the diffusion coefficient of the Brownian
motion that underlies the SLE and generates via a random sequence of
simple conformal maps the fractal curves. In addition to the fractal
dimension it determines various other geometric and statistical
properties of the SLE curves. One of them is for instance the
probability that a curve in the upper half plane $\mathbb{H}$ generated
by SLE will pass to the left of a given point $z=x+iy$ 
is given by Schramm's ``left passage formula'' \cite{schramm}
\be
P_\kappa(z)=\frac{1}{2}+\frac{\Gamma\left( \frac{4}{\kappa}\right) } 
{\sqrt{\pi}\,\Gamma\left( \frac{8-\kappa}{2\kappa}\right) }
\,\,_{2}F_{1}\left( \frac{1}{2},\frac{4}{\kappa};
\frac{3}{2};-\left(\frac{x}{y}\right)^2\right)\,\frac{x}{y}, 
\label{probphi}
\ee
where $_{2}F_{1}$ is the hyper-geometric function
$_{2}F_{1}\left( a_{1},a_{2};b;z\right) 
=\sum\limits_{k=0}\limits^\infty 
\frac{\Gamma\left(k+a_{1}\right)}{\Gamma\left(a_{1}\right)}
\frac{\Gamma\left(k+a_{2}\right)}{\Gamma\left(a_{2}\right)}
\frac{\Gamma\left(b\right)}{\Gamma\left(k+b\right)}
\frac{z^k}{k!}$.
Since the probability depends just on the ratio between ${\rm Re}(z)$ and
${\rm Im}(z)$, it is sometimes useful to replace this ratio by a function of
an angle.  We decided to use: $\tan{(\phi)}=x/y$. So
$\phi\in]-\frac{\pi}{2};\frac{\pi}{2}[$ is the angle at the origin
between the imaginary axis and $z$.  This formula holds for the domain
of the SLE being the upper half plane with the start point of the
curve being identical to the origin of the coordinate system and the
end point at infinity.

\begin{figure}[t]
\includegraphics[width=0.7\linewidth]{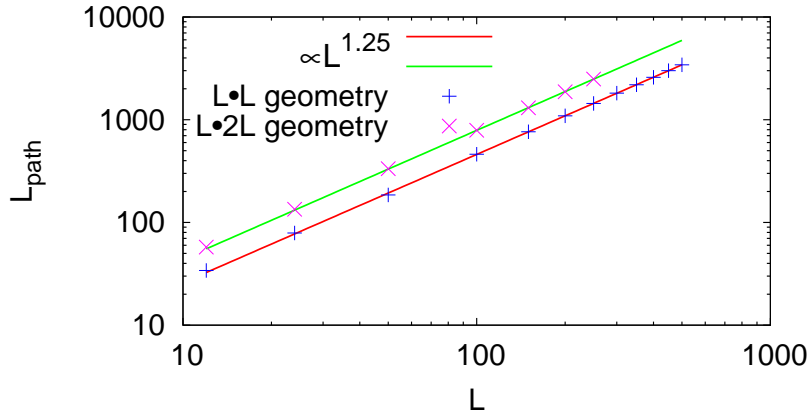} 
\caption{Average length of domain walls spanning the system from one 
end to the opposite one as a function of the system size $L$ in a
log-log plot. The straight lines are least square fits to (\ref{dwl})
and yield the estimate for the fractal dimension $d_s=1.25\pm0.01$.
}
\label{fig-frak}
\end{figure}

A standard check whether an ensemble of random curves is a potential
candidate for SLE therefore is to simultaneously determine their
fractal dimension $d_s$ and the left passage probabilities, and to
test whether the latter fit (\ref{probphi}) with $\kappa=8(d_s-1)$
\cite{amoruso,middleton-doussal,risaus-gusmann}. 

In Fig. \ref{fig-frak} we show our data for the average length 
of a domain wall in the configuration depicted in Fig.\ \ref{fig-dw2} b,
i.e.\ starting at the point $i_s=(1,1)$ and ending
at $i_t=(L,L)$ in a $L\times L$ geometry and 
at $i_t=(L,2L)$ in a $L\times 2L$ geometry. Least square fits 
to the scaling law (\ref{dwl}) yield the estimate 
$d_s=1.25\pm0.01$ which agrees with the value found in 
\cite{pfeiffer}. 

This value for the fractal dimension would imply $\kappa=2.00\pm0.08$
{\it if} the domain walls are described by SLE. Next we determined for
different points $(x,y)$ of the lattice the frequency that a domain
wall passes to the left of it, yielding a probability $p(x,y)$,
which we compared with Schramm's left passage formula $P_\kappa(x,y)$
for fixed $\kappa$.  For the circle domain and the choice
of the start and end points of the domain wall as shown in 
Fig.\ \ref{fig-dw2} c, the formula (\ref{probphi}) is modified: 

\begin{figure}[t]
\includegraphics[width=0.5\linewidth]{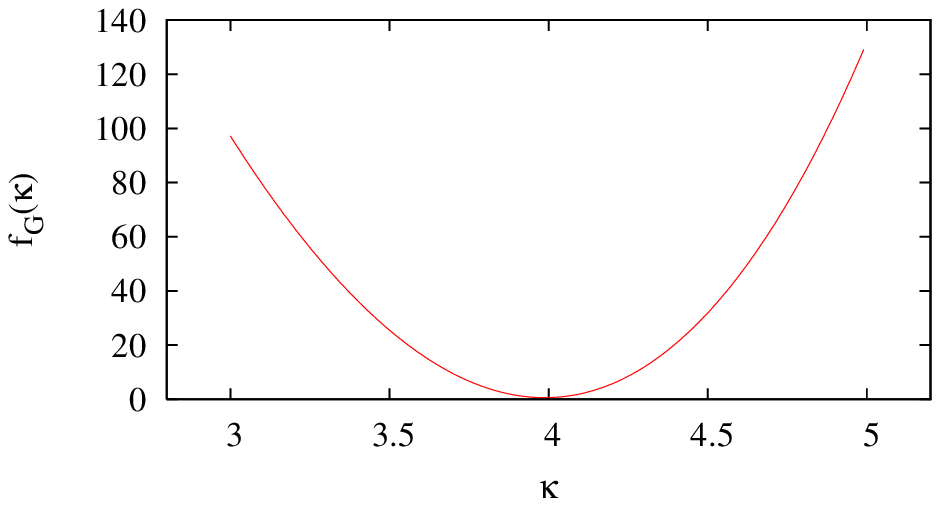} 
\includegraphics[width=0.5\linewidth]{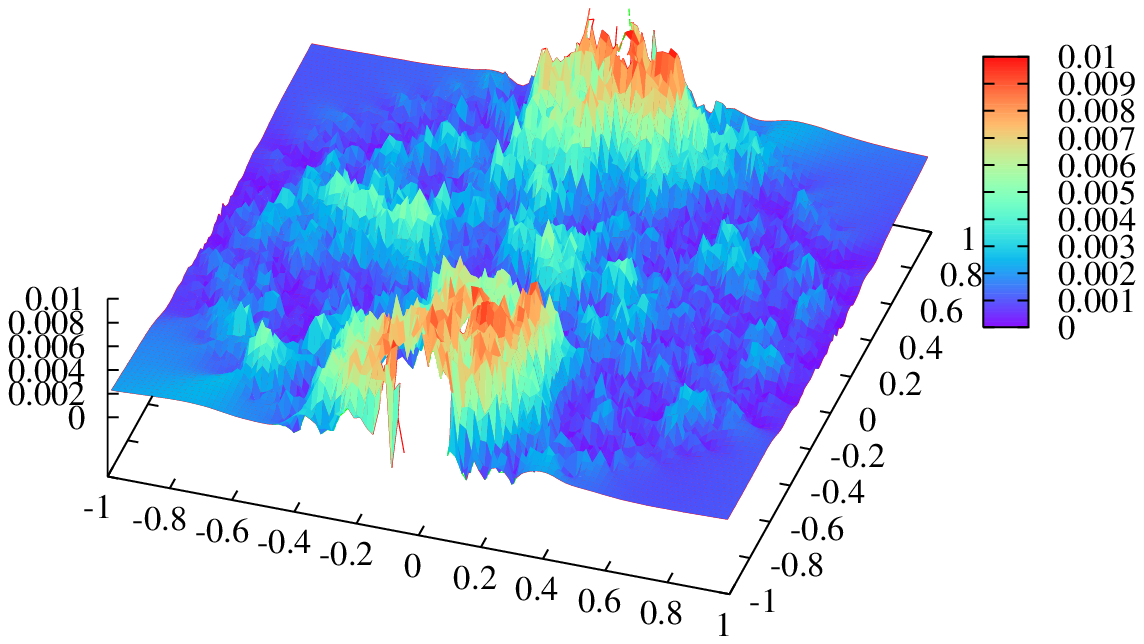} 
\caption{
{\bf Left}: The cumulative squared deviation $f_G(\kappa)$ of the computed
left passage probabilities $p(x,y)$ from the values
$P_{\kappa,g}(x,y)$ given by (\ref{gl2}) as a function of
$\kappa$. The underlying lattice geometry is the circle as sketched in
Fig. \ref{fig-dw2}c., the corresponding conformal map $g(z)$ entering
(\ref{gl2}) is given in the text. The minimum is at
$\kappa=4.00\pm0.01$ with a squared difference per grid point of about
$2\cdot10^{-5}$.  {\bf Right}: Absolute difference between the calculated
left passage probability $p(x,y)$ and the SLE expectation
$P_{\kappa,g}(x,y)$ (\ref{gl2}) for $\kappa=4$ as a function of the 2d
lattice coordinates $(x,y)$. For the whole unit circle the deviation
is almost everywhere smaller than 1\%.  Note that the domain wall is
fixed at $(-1,0)$ and $(1,0)$, where the largest deviations occur.}
\label{figKreis}
\end{figure}

\begin{figure}[t]
\includegraphics[width=0.5\linewidth]{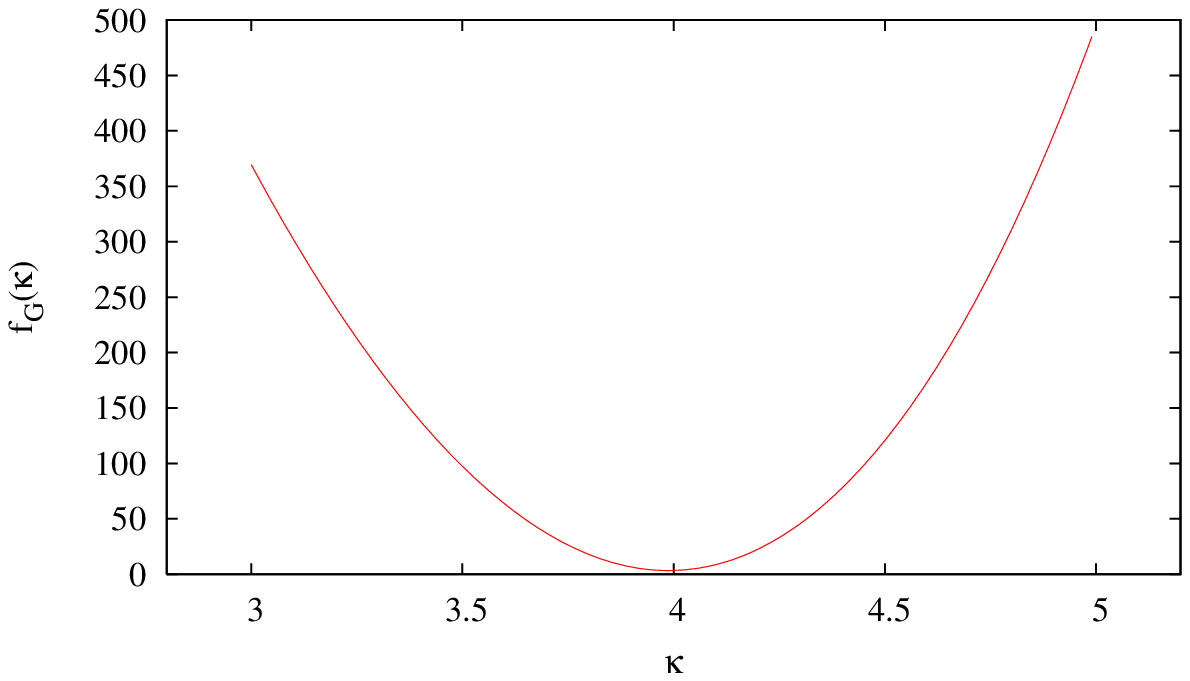} 
\includegraphics[width=0.5\linewidth]{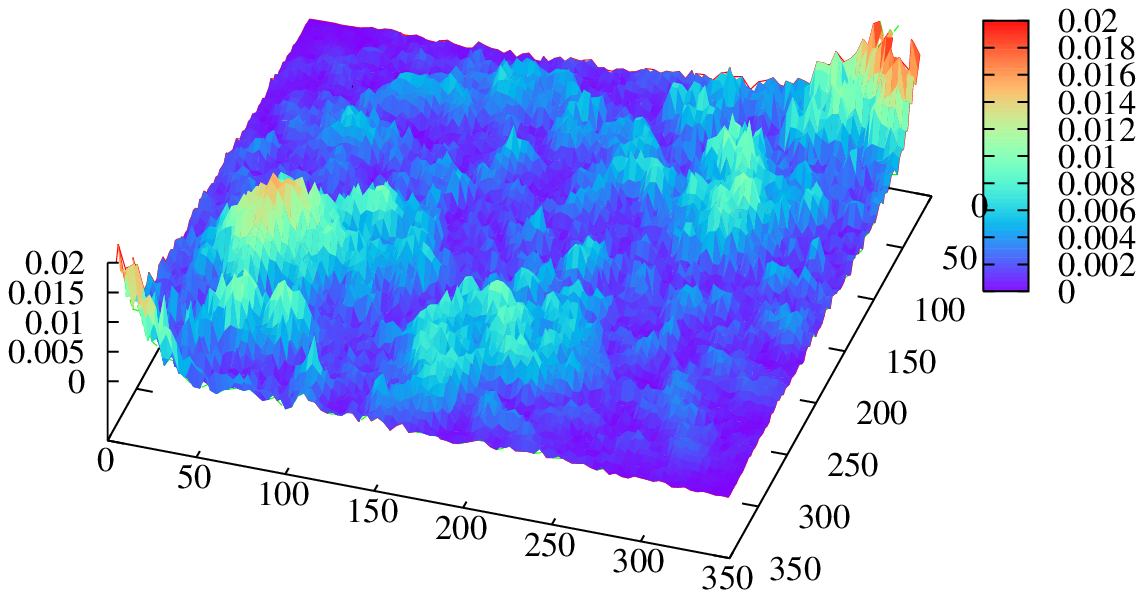} 
\caption{Cumulative deviation $f_G(\kappa)$ as function of $\kappa$
and absolute difference between $p(x,y)$ and $P_{\kappa,g}(x,y)$ for
$\kappa=4$ as in Fig.\ \ref{figKreis} but for the square geometry as
depicted in Fig.\ \ref{fig-dw2}b. Note that the domain wall is fixed
at $(0,0)$ and $(L,L)$, where the largest deviations occur.}
\label{figEcke}
\end{figure}

Let $\mathbb{E}$ be the unit circle in the complex plane.  The Cayley
function $g:\mathbb{E}\rightarrow\mathbb{H},z\mapsto
i\frac{1+z}{1-z}$ maps the unit circle conformally into the upper half
plane $\mathbb{H}$. Furthermore $g(-1)=0,\,g(1)=\infty$, thus curves
in $\overline{\mathbb{E}}$ starting at $z=-1$ and ending at $z=1$ are
conformally mapped on curves in $\overline{\mathbb{H}}$.  For the
latter, if the curves are described by SLE, Schramm's formula
(\ref{probphi}) holds, so that for the former, the modified formula
\be
P_{\kappa,g}(z)=
\frac{1}{2}+\frac{\Gamma\left( \frac{4}{\kappa}\right) } 
{\sqrt{\pi}\,\Gamma\left( \frac{8-\kappa}{2\kappa}\right) }\,\,
_{2}F_{1}\left( \frac{1}{2},\frac{4}{\kappa};\frac{3}{2};
-\left(\frac{{\rm Re}\,g(z)}{{\rm Im}\,g(z)}\right)^2 \right)\,
\frac{{\rm Re}\,g(z)}{{\rm Im}\,g(z)}.
\label{gl2}
\ee
holds. Hence we filled the unit circle with finer and finer grids $G$
approximating better and better the continuum limit for the curves in
$\mathbb{E}$ that we want to check for SLE. For this we define the
function
\be
f_G(\kappa)=\sum_{(x,y)\in G} [p(x,y)-P_{\kappa,g}(x,y)]^2
\ee
that measures the cumulative squared deviation of the probabilities $p(x,y)$
from the SLE-value for given $\kappa$. The result is shown in Fig.
\ref{figKreis}. The minimal deviation of the data from the 
expected SLE result is at $\kappa=4.00\pm0.01$ which clearly differs
from the value $\kappa=2.00\pm0.08$ that one would expect from the
fractal dimension if the domain walls would be described by SLE. 

Next we varied the geometry and the domain wall constraints and
studied the case depicted in Fig. \ref{fig-dw2}b, i.e.\ a quadratic
domain $D$ with corners at $0$, $p$, $p+ip$, $ip$ ($p$ real and
positive).  The function
$g:D\rightarrow\mathbb{H}:z\mapsto-\mathfrak{p}(z;S)$ with
$S=\left\lbrace 2 n_1 p+i\cdot 2 n_2 p|\,n_1,\,n_2 \in Z\right\rbrace$
defines a conformal map from $D$ into $\mathbb{H}$, where
$\mathfrak{p}(z;S)=\frac{1}{z^2}+
\sum\limits_{\omega\in S\setminus\left\lbrace0\right\rbrace} 
\left[ \frac{1}{\left( z-\omega\right)^2 }
- \frac{1}{\omega^2}\right]$ is the Weierstra\ss\ $\mathfrak
p$-function.  Furthermore $g(p+ip)=0,\,g(0)=\infty$, thus curves in
$\overline{D}$ starting at $z=0$ and ending at $z=p+ip$ are
conformally mapped on curves in $\overline{\mathbb{H}}$ starting at
the origin and extending to infinity. Using $g$ in (\ref{gl2}) we
determined $\kappa$ in the same way it was done in the case of the
unit circle. The result is shown in Fig. \ref{figEcke} and yields
$\kappa=4.00\pm0.01$.

\begin{figure}[t]
\includegraphics[width=0.8\linewidth]{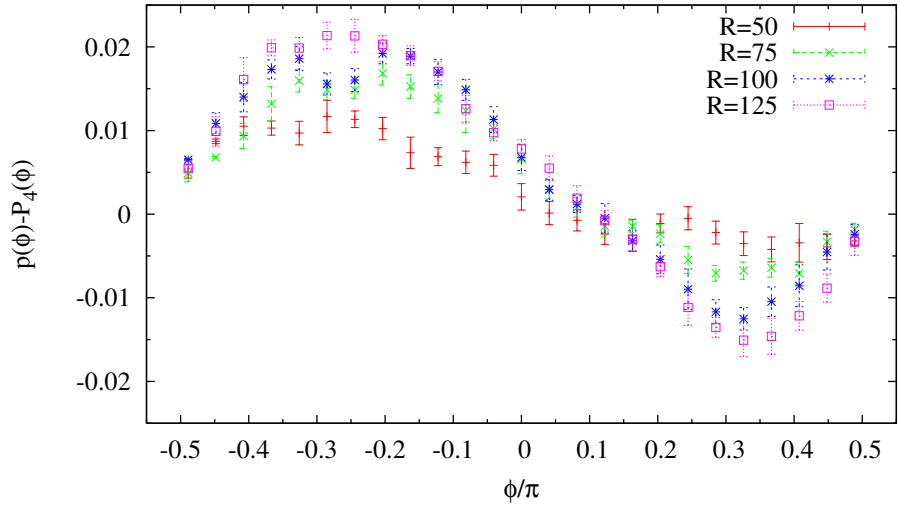} 
\caption{Deviation of the left passage probability $p(\phi)$ from 
the SLE expectation $P_\kappa(\phi)$ (\ref{probphi}) for $\kappa=4$ in
the half circle geometry depicted in Fig. \ref{fig-dw2} d
($p=510,\,q=255,\,R=500$), i.e.\ one free end of the domain wall.  }
\label{fig-phi}
\end{figure}

We also compared $p(x,y)$ directly with $P_{\kappa=4}(x,y)$ for curves
in the upper half plane $\mathbb{H}$ starting at the origin. For this we
considered the geometry depicted in Fig.\ \ref{fig-dw2}d, in
which the curves starting at the origin can end everywhere on the half
circle. We checked the prediction (\ref{probphi}) for different
radii $R$ and angles $\Phi=arctan\left( \frac{x}{y} \right)\,\in
\left] -\frac{\pi}{2}; \frac{\pi}{2} \right[$. The result is shown in
Fig. \ref{fig-phi}, where the deviations from $P_\kappa(\phi)$
(\ref{probphi}) for $\kappa=4$ are everywhere less than 2\%
for the size $L=500$ and radii shown in Fig.\ \ref{fig-phi}.
We also observe that the deviation systematically
decreases with the system size $L$ for fixed ratio $R/L$
indicating vanishing deviations $P(\Phi)-P_{\kappa=4}(\Phi)$
in the limit $L\to\infty$.

For all the geometries that we studied we found that the data can be
nicely fitted with $\kappa=4$, but the fractal dimensions of the domain
walls is unchanged by the different set-ups: $d_s=1.25\pm0.01$.  The
conclusion is that domain walls in the SOS model on a disordered
substrate are {\it not} described by SLE. The question arises which
condition for SLE is actually violated. The fact that for all
geometries the left passage probability we studied is well described
by a common expression, Schramm's formula containing the conformal map
of the half plane to the specific geometry under consideration, would
actually hint at conformal invariance. But obviously this does not
exclude the possibility that a breaking of conformal invariance 
manifests itself in other quantities, or even other geometries.
We did not attempt to check the domain Markov property, as 
was tried in~\cite{middleton-doussal}.

\section{Excitations}

In this section we study large scale excitations that cost a minimum
amount of energy, also denoted as droplets \cite{fisher-huse}, and the
size dependence of their energy. According to the usual arguments in
droplet scaling theory \cite{fisher-huse} this excitation energy is
expected to scale in the same way as a domain wall of lateral size
$L$, i.e.\ like $\Delta E\sim L^\theta$ with $\theta$ the stiffness
exponent, which for the system we consider is $\theta=0$. The domain
wall energy scales as $\Delta E_{\rm DW}\sim\log L$ c.f.\ (\ref{dwe}),
and if it is correct that all excitations of scale $L$ display the
same behavior the question arises: how can thermal fuctuations
destabilize the ground state such that the present model is indeed
characterized by a line of fixed points, indexed by temperature (and
thus different from a $T=0$ fixed point like in spin glasses with
$d>2$? A $\log L$ scaling of excitations still implies that larger and
larger excitations, necessary to occur to decorrelate the system's
state from the ground state, need more and more energy and are
therefore more and more unlikely, i.e.\ occurring with a probability
that decays as $\exp(-\Delta E/T)\sim L^{-c/T}$ at $T>0$.

\begin{figure}[t]
\includegraphics[width=0.8\linewidth]{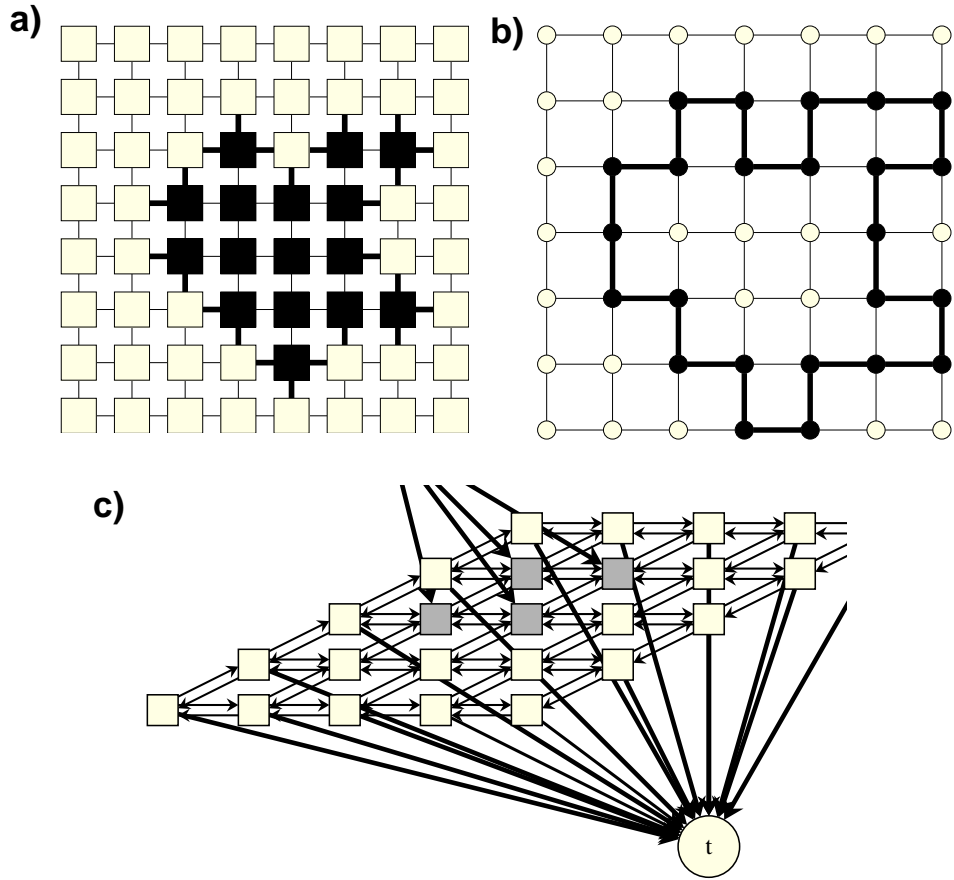} 
\caption{a) Sketch of a droplet excitation representing
a simply connected cluster ${\cal C}$ of sites $i$ whose height values
are increased by one as compared to the ground state: $n_i=n_i^0+1$
$\forall i\in {\cal C}$. b) Directed cycle in the dual lattice
corresponding to the cluster in a) representing the step in height
profile induced by raising the cluster ${\cal C}$ by one height unit. c)
Sketch of an $s-t$ cut of the nodes of the original lattice graph such
that those sites connected to the external node $s$ are forced to be
in the set $S$ and those connected to $t$ in the set $T$ (see text).}
\label{fig-exc-sketch}
\end{figure}

The solution to this problem lies in the freedom that domain walls of
excitations of scale $L$ have to optimize their energy (c.f.\ for a
similar discussion in the context of optimized dislocations in
\cite{pfeiffer}). Finding the optimal excitation of a given scale requires
a higher effort than searching a domain wall with given
start and endpoints (as we have seen above, there are even efficient
ways to optimize the positions of these endpoints).
There is a simple way to induce excitations of an {\it arbitrary}
scale, as first proposed in the context of spin glasses
\cite{kawashima} and used again in \cite{hartmann}:
we could compute the ground state for fixed boundary conditions as
depicted in the left panel of Fig.\ \ref{fig-dw}, this yields ${\bf
n}^0$. Then we choose a central site (say $i=(x,y)=(L/2,L/2)$), fix it
to $n_i=n_i^0+1$, fix the boundary as before and compute the ground
state again, giving ${\bf\tilde n}^0$. This will differ from ${\bf
n}^0$ only by a compact cluster $C$ that contains the site $i$ and
which has $\tilde n_i^0=n_i^0+1$. This is then in fact an optimal or
droplet excitation, but its size $V=\#\{i\in C\}$ can vary from 1 to
$L^2$. Droplets of a {\it fixed} size cannot be generated in this way.

The SOS model on a disordered substrate actually allows for
an efficient search for optimal excitations of a given scale,
as we describe in the following:

\begin{figure}[t]
\includegraphics[width=0.8\linewidth]{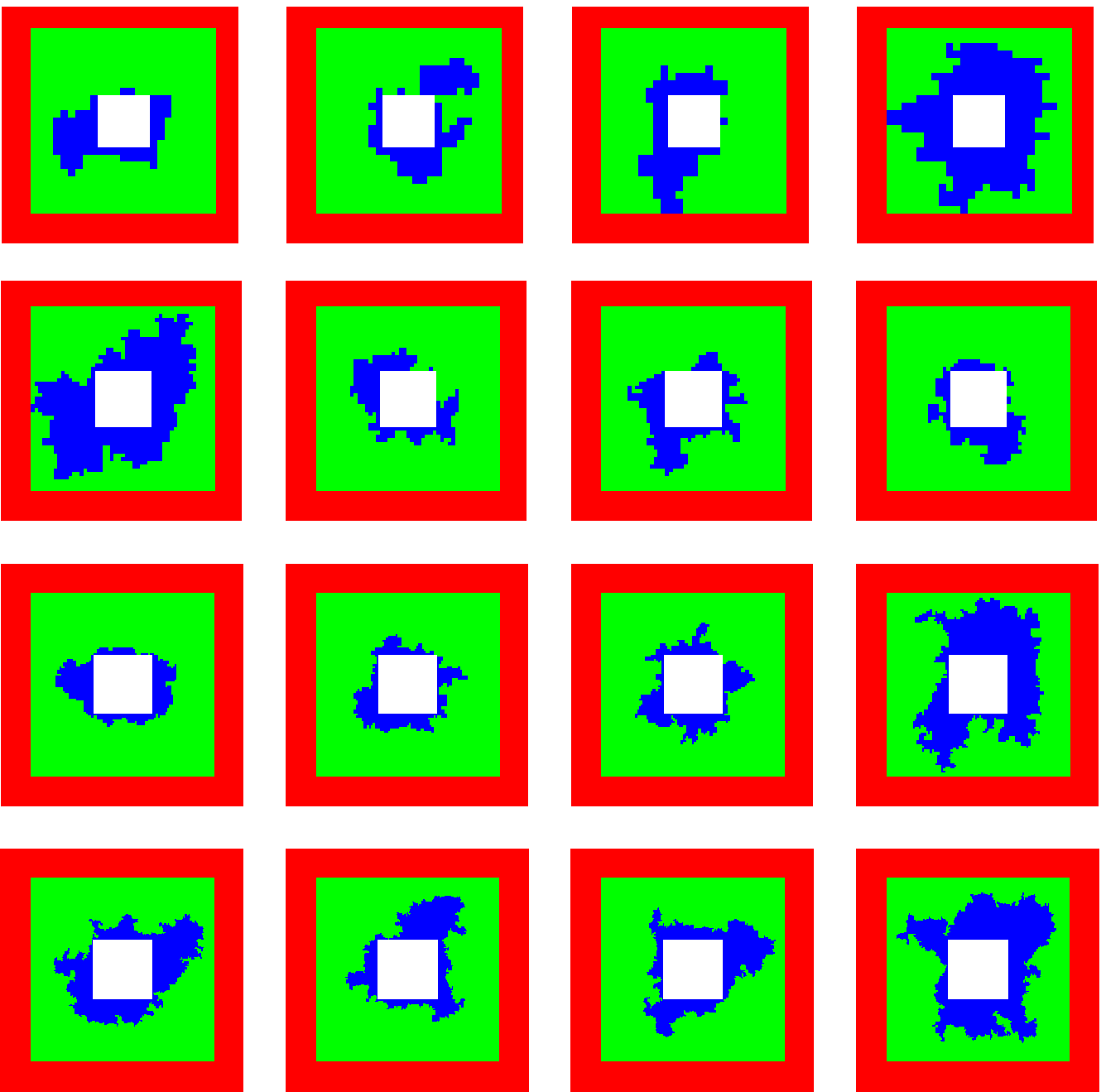} 
\caption{Examples of optimal excitations of scale $L$, whose 
boundary (domain wall) is forced to lie in the interior of the area
indicated in green (an annulus in square geometry). From the top row
to the bottom it is $L=32$, $64$, $128$ and $256$. Pictures are scaled
to have the same size.}
\label{fig-exc-examples}
\end{figure}

A droplet excitations in the SOS model is a simply connected cluster
${\cal C}$ of sites whose height values are all increased (or all
decreased) by one as compared to the ground state: $n_i=n_i^0+1$
$\forall i\in {\cal C}$, see Fig.\ \ref{fig-exc-sketch}. Pictorially
it is, in the height representation, an extra-mountain (valley) of
height $+1$ ($-1$). Its boundary is a directed cycle in the dual
lattice, and remembering the mapping to the minimum-cost-flow problem
(\ref{sos2}-\ref{con}), adding a cycle to a feasible solution
maintains the divergence-free constraint (\ref{con}). Hence, the
transition from one state to the other is determined by such a
cycle. Thus first one has to compute the ground state of a given
disorder realization.  Then in order to find a droplet
excitation of this ground state that has a given lateral size one can
for instance force this extra-cycle to run within an annulus of inner
radius $L/4$ and outer radius $3L/4$ (i.e\ it's average diameter is
$L/2$). This can be achieved by simply removing all sites / bonds
outside this annulus and then computing the optimal cycle within this
modified graph, the cost for which depends on the ground state
configuration and the substrate heights. One assigns costs to each
directed edge that corresponds to the energy cost for increasing the
height difference between its left and right side by one
unit~\cite{mincost,blasum,hartmann-rieger}. Note that in practice one
may use the reduced costs emerging from the ground state. That is, if
and only if a feasible flow has minimum energy, then there are
non-negative reduced costs such that the costs of each cycle remains
unchanged.  If the successive shortest path algorithm is used to solve
the minimum-cost flow-problem when computing the ground state ${\bf
n}^0$, then no extra work is necessary to compute the desired
non-negative reduced costs. Hence, Dijkstra's shortest path algorithm
can be used to find the shortest directed cycle around the annulus,
i.e.~the one separating the inner and the outer ring of the
annulus. To this end, the annulus is cut from the outer to the inner
ring, i.e.~the corresponding edges are removed from the graph, to
prevent the shortest path from short-cutting. For each of these removed
edges, the shortest path from its head to its tail is computed, where
only the the remaining edges are used. The obtained shortest paths are
completed with the corresponding directed edges to form cycles around
the annulus. This procedure has to be repeated for all (or a
representative number of) positions of the annulus within the original
lattice (in practice one fixes the annulus and shifts the disorder
configurations, wrapping it around a toroidal geometry).

\begin{figure}[t]
\includegraphics[width=\linewidth]{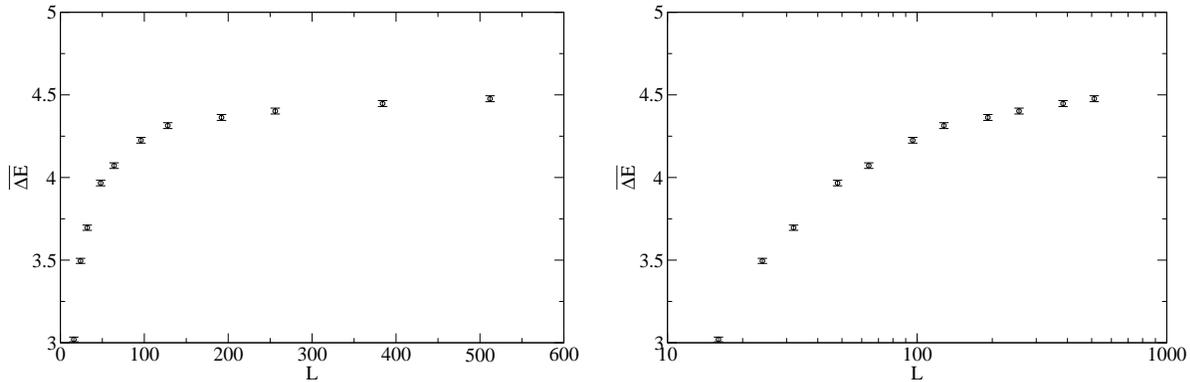} 
\caption{Disorder averaged energy of the optimal excitations
of scale $L$ as a function of $L$ - left in linear scale, right in a
log-log scale.}
\label{fig-exc-dat}
\end{figure}

The procedure of finding the optimal cycle in a given annulus can be
simplified by observing that the droplet excitation in the height
representation corresponds to an $s$-$t$-cut of the underlying graph
\cite{hartmann-rieger}
\footnote{An $s$-$t$-cut is a partition of the
nodes of a graph ${\cal G}$ into two disjoint sets ${\cal S}$ and
${\cal T}={\cal G}/{\cal T}$, such that $s\in S$ and $t\in T$.} in
such a way that one forces all nodes of the inner circle of the
annulus to belong to $S$ and all nodes belonging to the outer ring of
the annulus to $T$, see Fig.\
\ref{fig-exc-sketch} c. 
The minimum $s$-$t$-cut with respect to the non-negative reduced
edge costs of the ground state ${\bf n}^0$ is then exactly the 
boundary of the optimal excitation 
(or the optimal cycle) one is searching for a given annulus
arrangement.
According to the famous Min-Cut-Max-Flow theorem 
one can compute the minimum $s$-$t$ cut in polynomial time
by solving the associated maximum-flow problem
\cite{blasum,hartmann-rieger}.
We have implemented this procedure and show for illustration 
a number of examples in Fig.\ \ref{fig-exc-examples}.

Fig.\ \ref{fig-exc-dat} shows our result for the disorder averaged
energy of the optimal excitations of scale $L$ that we obtain with the
procedure described above. This represents an upper bound for the
optimal excitations of scale $L$ since the annulus arrangement does
not include all possible excitations of scale $L$. As one can see this
bound saturates at a finite energy of order ${\cal O}(1)$ in the limit
$L\to\infty$. Consequently arbitrarily large excitations exist that cost
only a small amount of energy, which renders the ground
state unstable at an non-vanishing temperature.

\begin{figure}[h]
\includegraphics[width=0.5\linewidth]{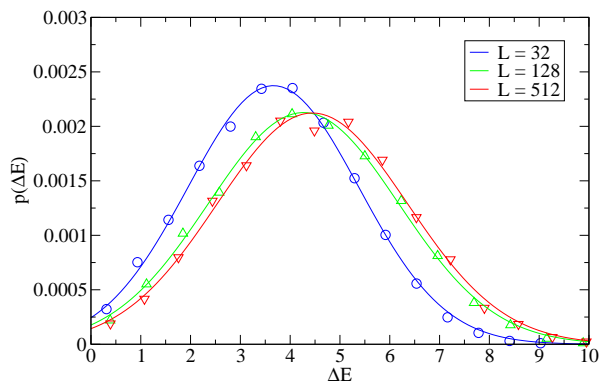} 
\caption{Distribution of the optimal excitation energies of scale $L$
for three values of $L$. For $L\ge 128$ this distribution becomes 
independent of $L$.}
\label{fig-exc-distr}
\end{figure}

Fig.\ \ref{fig-exc-distr} shows the distribution of optimal excitation
energies for different values of $L$. As can be seen the distribution
is nearly Gaussian with a finite width, i.e.\ it does not display
long, e.g.\ algebraically decaying, tails, which implies that the
average is representative for almost all disorder configurations. It
can also be seen that the whole probability distribution $P_L(\Delta
E)$ becomes independent of the length scale $L$ for large $L$. This is
an important observation since droplet scaling theory
\cite{fisher-huse} one
would expect $P_L(\Delta E)\sim l^{-\theta}\tilde{p}(\Delta E/L^\theta)$. For
$\theta=0$, as it is the case here, it is not a priori clear whether
this implies $P_L(\Delta E)\sim (\ln L)^{-1}\tilde{p}(\Delta
E/\ln L)$, i.e.\ a scaling with the average domain wall energy $\ln
L$, or $P_L(\Delta E)\sim \tilde{p}(\Delta E)$, i.e.\
droplet size independence.
Fig.\ \ref{fig-exc-distr} shows that the latter is correct.

This has important implications for the scaling of the average droplet
energy within a system of lateral size $L$ as determined in
\cite{middleton-drop}. There the average energy of droplets of size $l<L$,
$(\Delta E)_L^{\rm min}$, i.e.\ without a lower bound, was estimated
to behave like $(\Delta E)_L^{\rm min}\sim\ln L$. In order to make
contact with our result one should note that this energy is the
minimum among droplets of the kind we determined here, with size
$l\in[L/2,L]$, $l\in[L/4,L/2]$, $l\in[L/8,L/4]$, $\ldots$, i.e a
minmum of approximately $\ln L$ random numbers. Only if the
probability distribution of these energies is a) independent and b)
identically distributed, their minimum goes like the inverse of their
number, i.e.\ $(\Delta E)_L^{\rm min}={\rm min}\{(\Delta E)_{L/2^1},
(\Delta E)_{L/2^2},\ldots, (\Delta E)_{L/2^k}\}\sim 1/k$ with
$k\sim\ln L$. According to our result depicted in Fig.\
\ref{fig-exc-distr} the assumption b, which is implicit in the
reasoning of \cite{middleton-drop}, is indeed fulfilled.

Finally we note that also determined the fractal dimension of the
boundary of the optimal excitations and found that it is identical
with the fractal dimension of the domain walls with fixed start and
endpoints: $d_s=1.25\pm0.01$.

\section{Disorder Chaos}

In this section, we study 
the sensibility of the ground state to a small change
of the quenched disorder configuration. To this purpose we generate a
configuration of random offsets $d^1_i$ and compute the associated ground state
$h_i^1$. Then we slightly perturb this configuration
of random offsets $d^2_i = d_i + \delta \epsilon_i$ with $\delta \ll 1$ and
where $\epsilon_i$'s are independent and identically distributed
Gaussian variables of unit 
variance and we compute the associated ground state $h_i^2$. The
question we ask is : how different are these two configurations of the
systems $h_i^1$ and $h_i^2$ ?

Such questions first arose in the
context of spin glasses \cite{bray-moore}, where it was proposed
that disorder induced glass phases may exhibit ''static chaos'', {\it
  i.e.} extreme 
sensitivity to such small modifications of external parameters (like
disorder considered here or temperature). Such small perturbations are
argued to decorrelate 
the system beyond the so called {\it overlap length} $L_\delta$ which
diverges for small $\delta$ as $L_\delta \sim \delta^{-1/\alpha}$,
with $\alpha$ the chaos exponent. As an example let us first consider the case of an Ising
spin-glass with a continuous Gaussian distribution of
random exchange interactions, of width $J$. The system has two ground states related
by a global spin reversal. Within the phenomenological droplet theory
\cite{bray-moore, fisher_droplet, florent_bouchaud}, a low-lying energy
excitations of the system involves an overturned droplet of linear
size $L$ and costs an energy $J {L }^\theta$. If we now add a
small random bond perturbation, say Gaussian of width $\delta J$, the
excess energy of a droplet is modified. For such spin-glass system, this energy comes
only from the bonds which are at the {\it surface} of the droplet. Their
contribution is thus the sum of ${L}^{d_s}$ independent random
variables of width $\delta J$, where $d_s$ is the fractal dimension of
the droplet : it is thus of order $\pm \delta {\ell}^{d_s/2}$. Therefore the
ground state is unstable to the perturbation on length scales $L$ such that $\delta J  L^{d_s/2} > J L^\theta$, {\it i.e.} 
$L > L_\delta$ where $L_\delta \sim \delta^{-1/\alpha_{\rm SG}}$ with
$\alpha_{\rm SG} = d_s/2 - \theta$. One thus sees that, for spin
glasses, disorder chaos is closely related to the (geometrical)
properties of the domain walls. 

The situation is rather different for elastic systems in a random
potential as considered here. Here we consider the SOS model on a disordered substrate defined in Eq. (\ref{sos}) as
\begin{eqnarray}\label{sos_chaos}
H = \sum_{(ij)} (h_i-h_j)^2 \; , \; h_i = n_i +d_i \;,
\end{eqnarray}
with $i \equiv (x_i,y_i) \in {\mathbb Z}^2$. In Eq. (\ref{sos_chaos}) the height variables $n_i$ ($i=1,\ldots,N$) take on integer
values $n_i=0,\pm1,\pm2,\ldots$ and the offsets $d_i$ are independent quenched
random variables uniformly distributed between 0 and 1. This system
(with free of periodic 
boundary conditions) has infinitely many ground states which differ by a global shift $\Delta n \in \{\pm 1, \pm 2,... \}$. Again,
within the droplet argument \cite{fisher_droplet, shapir_chaos, pld_chaos} a
low-lying energy excitation of the system involves a droplet of size
$L$ where the height field is shifted by unity, say $\Delta n = 1$
and it costs an energy ${L }^\theta$.  But now if we had a small
perturbation $d^2_i = d^1_i + \delta \epsilon_i$, the excess energy of
this droplet comes from the {\it bulk} of the droplet, which
experiences a random force field. This contribution is thus the sum of
${L}^d$ random variables of width $\delta$ and therefore in this
case the ground state is unstable to the perturbation on length scales ${L}
> L_\delta$ where $L_\delta \sim \delta^{-1/\alpha}$ with
$\alpha = d/2 - \theta$. Here $d=2$ and $\theta = 0$ and thus one expects
$L_\delta \propto \delta^{-1}$. At variance with spin-glasses
discussed above, one thus sees that for disordered
elastic systems, disorder chaos is not directly related to the
properties of domain walls.

For the present model (\ref{sos_chaos}), disorder chaos was demonstrated analytically at finite $T$ near the glass transition
$T_g$ using a Coulomb Gas Renormalization Group
\cite{hwa_fisher}. At $T=0$, some indications of disorder
chaos were also found numerically in Ref. \cite{blasum} where global correlations
between $h_i^1$ and $h_i^2$ where studied through $\chi(\delta) =
\sum_{i} (h_i^1 - h_i^2)^2$. In this paper, following Ref. \cite{pld_chaos}, we
characterize the local correlations between these two configurations by the
correlation function $C_{ij}({\bf r})$ with ${\bf r} \equiv (x,y)$
(with $i,j = 1,2$)  
\begin{eqnarray}\label{def_correl_chaos}
C_{ij}({\bf r}) = \overline{(h_k^i - h^i_{k+{\bf r}})(h_k^j -
  h^j_{k+{\bf r}})} \; ,
\end{eqnarray}
where $k+{\bf r} \equiv (x_k + x, y_k+y)$ and for a rotationnaly
invariant system considered here one has $C_{ij}({\bf r}) \equiv C_{ij}(r)$
with $r = |{\bf r}|$. In the following we will talk about {\it
  intralayer} correlations for $C_{ii}(r)$ 
and {\it interlayer} correlations for $C_{i \neq j}(r)$. Equivalently,
one can also study such correlations (\ref{def_correl_chaos}) in
Fourier space and define $S_{ij}({\bf q})$ with ${\bf q} \equiv (q_x, q_y)$ as:
\begin{eqnarray}\label{struct_fact}
S_{ij}({\bf q}) = \overline{\hat h_{q}^i \hat h_{-{q}}^j} \; , \; \hat
h_{q}^j = \frac{1}{L^2} \sum_k h_k^j e^{i {\bf q} \cdot k} 
\end{eqnarray}
where ${\bf q} \cdot k = q_x x_k + q_y y_k$. For a rotationnaly
invariant system
considered here, one has $S_{ij}({\bf q}) \equiv S_{ij}({q})$ where $q =
|{\bf q}|$.  

Disorder chaos at $T=0$ in generic disordered elastic systems in
dimension $d$ was recently studied analytically using the 
Functional Renormalization Group (FRG) \cite{pld_chaos}. At one loop order in
a dimensional expansion in $d = 4 - \epsilon$ it was found that for short range disorder (like random bond problems) and random periodic
systems in dimension $ d > 2$ 
(including one component Bragg-Glass), one has \cite{pld_chaos}
\begin{eqnarray}\label{frg_predict}
C_{12}(r) = r^{2 \zeta} \Phi(\delta r^\alpha) \quad {\rm with} \quad
\Phi(x) \sim
\cases{
c^{\rm st} \quad, \quad x \ll 1 \;,\\
x^{-\mu} \quad, \quad x \gg 1 \;,
}
\end{eqnarray}
with $c^{\rm st}$ a constant, $\zeta$ the roughness exponent
and where $\mu$ is the {\it decorrelation} exponent. Translated into
Fourier space, this yields 
\begin{eqnarray}\label{struct_frg}
S_{12}(q) = L_{\delta}^{(d+2\zeta)} \varphi(q L_{\delta}) \quad {\rm
  with} \quad 
\varphi(x) \sim
\cases{
x^{-d-2\zeta+\mu} \quad, \quad x \ll 1 \;,\\
x^{-d-2\zeta} \quad, \quad x \gg 1 \;,
}
\end{eqnarray}
with $L_\delta \sim \delta^{-1/\alpha}$ and where the behavior for
large $x$ is then such that the dependence on $L_\delta$ cancels in
this limit, as it should.

The two-dimensional disordered SOS model we are considering here (\ref{sos_chaos})
corresponds precisely to the marginal case $d=2$ (with $\zeta = 0$ and
thus $\theta = 0$) where, as discussed in Ref. \cite{pld_chaos,
  duemmer_chaos}, the analysis yielding the result in
Eq. (\ref{frg_predict}) ceases to 
be valid. Indeed in that case non
local terms, irrelevant in $d > 2$ are generated under coarse-graining
and these additional terms have to be handled with care at $T=0$. One thus considers the Hamiltonian associated
to the two copies of the system parametrized by the scalar fields
$u^{i} \equiv u^{i}(\bf r)$:
\begin{eqnarray}\label{h_2copies}
H_{2 \, {\rm copies}} = \frac{1}{T} \sum_{i=1,2} \int d^2 {\bf r} \left[  \frac{1}{2}(\nabla_{\bf r} u^i)^2 + V_i(u^i,{\bf r}) - {\boldsymbol \mu}^i({\bf r}) \cdot \nabla_{\bf r} u^i \right] \;,
\end{eqnarray}
where ${\bm \mu}^{i} \equiv
(\mu_x^i, \mu_y^i)$ are two-component random tilt fields, which are
generated upon renormalization. While they are irrelevant in $d>2$
they become relevant in $d=2$ where they play a crucial role. These
two copies $u^{1}, u^2$ are thus independent (\ref{h_2copies}) but
they feel two mutually correlated random potentials 
\begin{eqnarray} 
&&\overline{V_i(u,{\bf r}) V_j(u',{\bf r'})} = R_{ij}(u-u') \delta^2({\bf r}-{\bf r'}) \;, \\
&&\overline{\mu_\rho^i({\bf r}) \mu_{\rho '}^j({\bf r'}) } = \sigma_{ij} \delta_{\rho \rho'} \delta^2({\bf r}-{\bf r'})  \;,
\end{eqnarray}
where $i = 1,2$ is the index of the copy and $\rho = x,y$ is a spatial index. This leads to the replicated Hamiltonian, $\overline {Z^n} = \exp{(-H_{\rm rep})}$:

\begin{eqnarray}\label{frg_oneloopnonloc} 
H_{\rm rep} = \frac{1}{2T} \int d^2 {\bf r} \sum_{i=1}^2 \sum_{a=1}^n
\frac{1}{2}(\nabla_{\bf r} u_a^i)^2 - && \frac{1}{2T^2}
\sum_{i,j=1}^2 \sum_{a,b=1}^n\int d^2{\bf r}
[  R_{ij}(u^i_a-u^j_b) \\
&& - \frac{1}{2} \nabla_{\bf r} u^{i}_{ab} \nabla_{\bf r}
u^{j}_{ab} G_{ij}(u^i_a - u^j_b) ] \nonumber \;,
\end{eqnarray} 
where we used the notation $u^{i}_{ab} = u^i_a - u^i_b$. In the
``bare'' model, one has $G_{ij}(u) = \sigma_{ij}$.   

Close to the transition $T \lesssim T_g$, this model
(\ref{frg_oneloopnonloc}) was studied using Wilson RG analysis
by varying the short scale momentum cutoff $\Lambda_{\ell} = \Lambda
e^{-\ell}$, with $\ell$ the log-scale. It was shown, using Coulomb Gas
technique at lowest order, that $G_{ii} \propto \sigma \ell$. This
leads to the correlation function in Fourier space $S_{ii}(q) \propto
\sigma \ell/{q}^2$, which yields, setting $\ell = \log(1/q)$, to the
$\log^2{(r)}$ behavior of the intralayer correlations (this result for
$C_{ii}(r)$ can be derived in a more controlled way using the Exact
Renormalization Group \cite{schehr_co}). Concerning chaos properties,
it was shown  
in Ref. \cite{hwa_fisher} that $G_{12}(0)$ grows linearly with 
${\ell}$ for small ${\ell}$ before it saturates to a constant for
large ${\ell}$, $G_{12}(0) \sim \hat \sigma > 0$, which yields
$S_{12}(q)\propto \hat \sigma/{q}^2$. These results close to $T_g$ can be
summarized as 
\begin{eqnarray}\label{struct_tg}
S_{12}(q) \sim \cases{
\sigma \frac{\log{1/q}}{q^2} \; ,\; q \gg L_\delta^{-1} \\
\frac{\hat \sigma}{q^2} \;, \; q \ll L_\delta^{-1}
}
\end{eqnarray}
while in real space, the behavior of the interlayer correlation function $C_{12}(r)$ for $T \lesssim T_g$ is thus
\begin{eqnarray}\label{chaos_tg}
C_{12}(r) \sim \cases{
\sigma \log^2(r) \quad, \quad r \ll L_\delta \\
\hat \sigma \log(r) \quad, \quad r \gg L_\delta \;.
}
\end{eqnarray} 

At $T=0$, it was recently shown \cite{doussal-schehr}, using FRG to
one loop including the term $G_{11}(u)$ that  
the intralayer correlation function also behaves like $C(r) \propto \log^2(r)$, in rather
good agreement with numerics \cite{blasum, zeng-sos}. One thus also
expects that 
$C_{12}(r) \propto \log^2(r)$ for $r 
\ll L_\delta$ \cite{blasum, zeng-sos,doussal-schehr}. For $r \gg
L_\delta$, a behavior of
$C_{12}(r) \sim \hat \sigma \log{r}$ as in Eq. (\ref{struct_tg}) was
discussed in Ref. \cite{pld_chaos}. To determine analytically whether
$\hat \sigma > 0$ at $T=0$ requires a
detailed and difficult analysis of the coupled FRG equations for
$R_{ij}(u), G_{ij}(u)$ (\ref{frg_oneloopnonloc}), which goes beyond
the previous studies done in that direction in
Ref. \cite{doussal-schehr, pld_chaos, duemmer_chaos}. Here, we will 
answer this question using numerical simulations.   

The purpose of our study is actually to answer the two main questions : (i) what are the residual correlations beyond $L_\delta$ and in particular is $\hat \sigma$ also finite at $T=0$~? (ii)  what is the scaling form of this correlation function $C_{12}(r)$, {\it i.e.} the
analogous of Eq. (\ref{frg_predict}) from which one can extract the
overlap length $L_\delta$ and check the value of chaos exponent
$\alpha = 1$ as expected from droplet scaling~?   

\begin{figure}
\begin{minipage}{0.5\linewidth}
\begin{center}
\includegraphics[angle = -90, width=\linewidth]{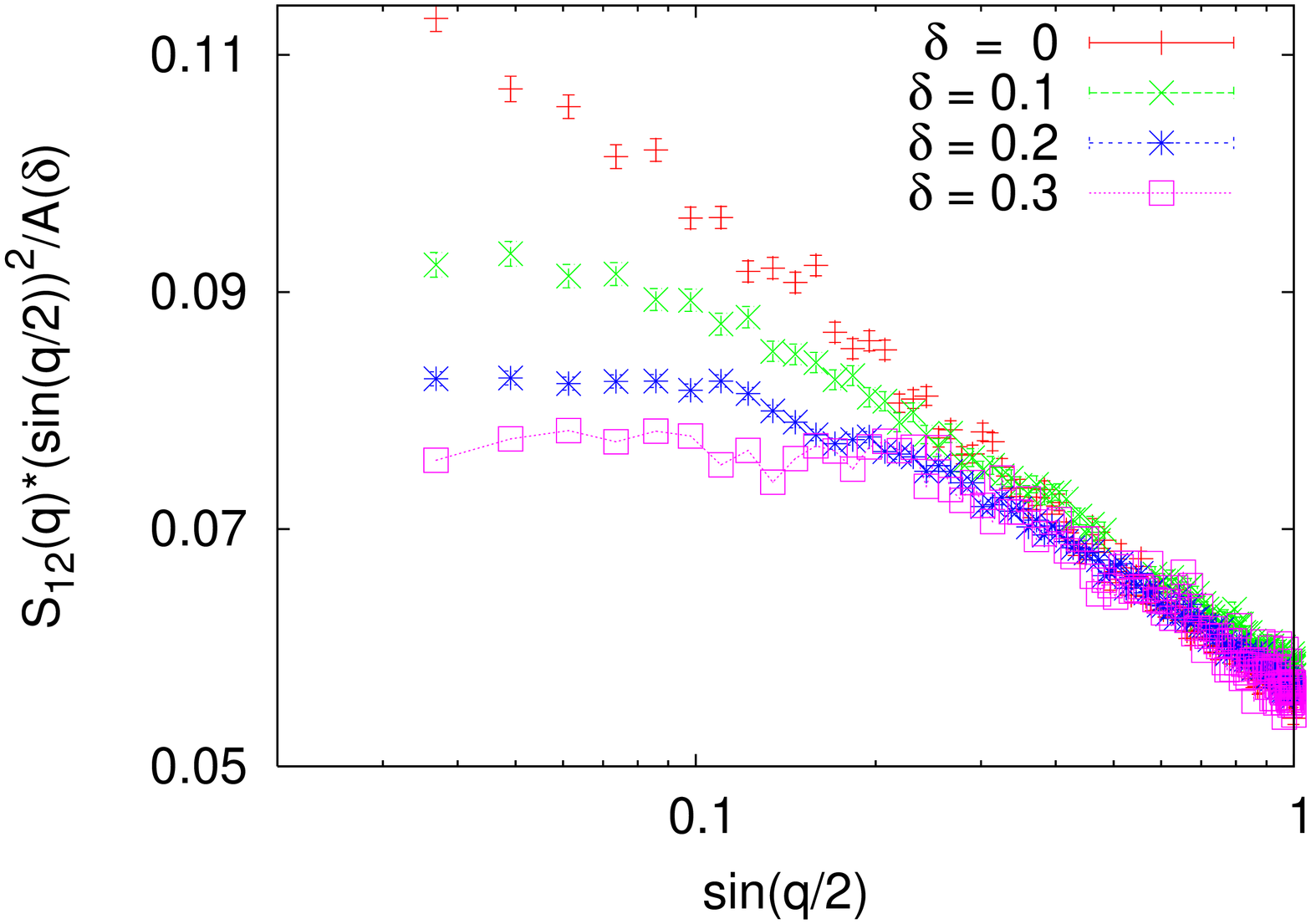}
\end{center}
\end{minipage}\hfill
\begin{minipage}{0.5\linewidth}
\begin{center}
\includegraphics[angle = -90, width=\linewidth]{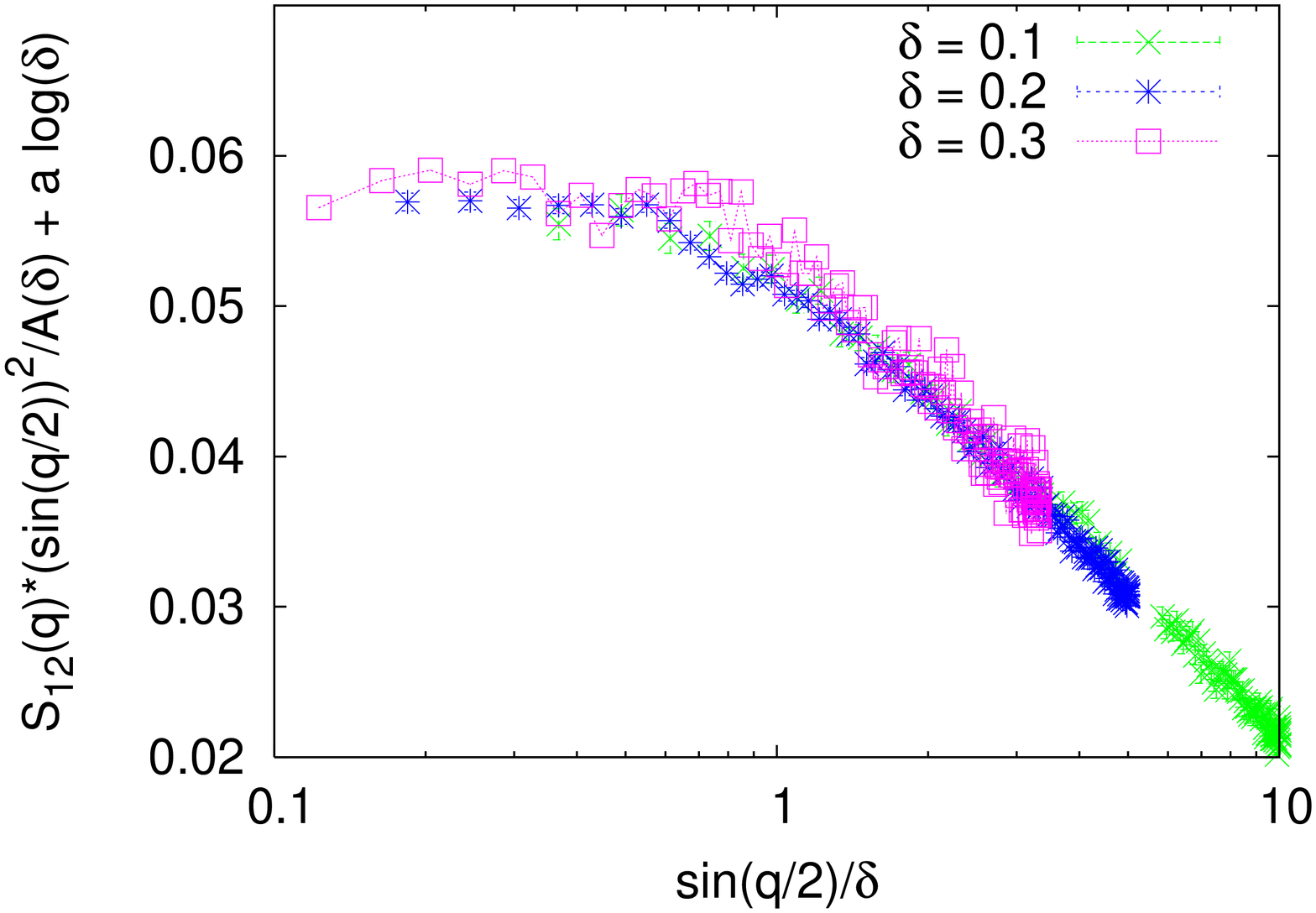}
\end{center}
\end{minipage}
\caption{{\bf Left : } Plot of $S_{12}(q) * y^2/A(\delta)$
  as a function of $ y = \sin{(q/2)}$ for different values of
  $\delta = 0, 0.1,0.2, 0.3$ for a linear system size $L = 256$ on a
  log-linear scale. The deviation from the straight line for $\delta >
  0$ is a clear indication of disorder chaos. This behavior is
  consistent with the one in 
  Eq. (\ref{struct_tg}) and in particular with a finite value of $\hat
  \sigma$ at $T=0$. {\bf Right :} Same data as the one shown in
  the right panel where we plot $y^2
S_{12}(q)/A(\delta)+ a\log{\delta}$ as a function of $y/\delta$. The
  relatively good collapse is consitent with the scaling form proposed
  in Eq. (\ref{scaling}), implying in particular $\alpha = 1$.}   \label{fig_chaos} 
\end{figure}

Here will use the minimum-cost-flow algorithm described above to
compute the two ground states $h_i^1$ and $h_i^2$ with free boundary
conditions. Instead of the correlation function $C_{12}(r)$ we compute
numerically the Fourier transform $S_{12}(q)$ of the ''overlap''
between the configurations (\ref{struct_fact}). Our simulations have
been performed on a square lattice of linear size $L=256$ and we have
chosen  
${\bf q} = (q,0)$ with $q = 2 \pi n/L$ with $n = 0, 1, 2, \cdots,
L-1$. The disorder average has been performed over $10^6$ independent
samples. 
In Fig. \ref{fig_chaos} (left) we show our numerical data for
$S_{12}(q)$. Its behavior close to $T_g$ (\ref{struct_tg}) suggests
strongly to plot $q^2 S_{12}(q)$ as a function of $q$. Given that we
are working on a discrete lattice of finite size, it is more
convenient to work with the variable $y = {\sin{(q/2)}} = [(1-
  \cos{(q)})/2]^{1/2}$ instead of $q$ (of course for small $q$ it
makes no difference). In Fig. \ref{fig_chaos}
(left) we actually show a plot of $S_{12}(q)[\sin{(q/2)}]^2/A(\delta)$
as a function of ${\sin{(q/2)}}$ on a log-linear plot for different
values of $\delta = 0,0.1,0.2,0.3$. On this plot the amplitude $A(\delta)$ is
chosen such that the curves for different values of $\delta$ do
coincide for ${\sin{(q/2)}} \sim 1$, with $A(0) = 1$. 
For $\delta = 0$, where $S_{12}(q)=S_{11}(q)$, this plot is almost a
straight line, which suggests indeed  that for small $q$ where
${\sin{(q/2)}} \sim q/2$,   
$S_{11}(q) \sim \log{(1/q)}/q^2$: this produces the $\log^2(r)$
behavior of the correlation function $C_{11}(r)$ in real space. This
result is consistent with previous numerical studies of the ground
state \footnote{In addition, this gives an alternative way to estimate the
amplitude of the $\log^2(r)$ term, here $a_2 = 0.40$ which is slightly
different from the previous estimate obtained from the fit of
$C_{11}(r)$ in real space and yielded $a_2 = 0.57$ (which actually
recovered in the present simulations by applying the same fitting
procedure).}.    

For $\delta > 0$ the behavior is however quite different : indeed for
small $q$, the curves deviate from the straight line indicating a
saturation to a finite value. We have checked that this is not a
finite size effect: in particular, for a given value of $\delta$, we
have checked that the value of $q$ where the bending occurs and the
saturating value do not depend on $L$. This bending is thus a clear
signature of 
disorder chaos in this model. In addition, the saturation to a finite
value is consistent value of $\hat \sigma$ at $T=0$. If one translates
these results in real space, one obtains a behavior of the correlation
function $C_{12}(r) \propto \hat \sigma \log{r}$ as in
Eq. (\ref{chaos_tg}). Of course $A(\delta) 
\to 1$ when $\delta \to 0$ but we were not able to characterize
precisely the dependence of $A(\delta)$. We emphasize that the
observation of disorder chaos here 
needs a precise computation of $q^2 S_{12}(q)$ and thus needs accurate
statistics.    

The scaling form in Eq. (\ref{struct_frg}) would suggest to plot $q^2
S_{12}(q)$ as a function of $L^{2 \zeta} \tilde \varphi(q
L_\delta)$. Here one has $\zeta = 0$ together with $\alpha = 1$
expected from scaling argument and on the other hand one
expects, see Fig. \ref{fig_chaos} (left) and also Eq. (\ref{struct_tg}), that
$\tilde \varphi(x) \sim - a \log{(x)}$ for large $x$ (with $a =
0.016(1)$ estimated from $q^2 S_{11}(q)$). Therefore, to  
guarantee that $S_{12}(q)$ has a good limit when $\delta \to 0$, we
propose the scaling form  
\begin{eqnarray}\label{scaling}
S_{12}(q) \sim q^{-2} \left( \tilde \varphi\left(\frac{q}{\delta}
\right) - a \log{\delta}   \right) \;, 
\end{eqnarray}
where the scaling function $\tilde \varphi(z)$ behaves like
\begin{eqnarray}
\tilde \varphi(z) = \cases{
c^{\rm st} \;, \; z \to 0 \\
- a \log{z} \;, \; z \to \infty 
}
\end{eqnarray}
such that the additional constant $-a \log{\delta}$ with $a =
0.016(1)$ (independently of $\delta$) is needed to yield a well
defined limit $\delta \to 0$. We 
have checked this scaling form (\ref{scaling}) for different
values of $\delta = 0.1, 0.2, 0.3$. Again, to take into account finite
size effects, we use the variable $y = {\sin{(q/2)}}$ instead of
$q$. In Fig. \ref{fig_chaos} (right) we show a plot of $y^2
S_{12}(q)/A(\delta)+ a\log{\delta}$ as a function of $y/\delta$. The
relative good 
collapse of these curves corresponding to different values of $\delta$
is in rather good agreement with the scaling in
Eq. (\ref{scaling}). This indicates in particular that $\alpha \simeq
1$, although a study of $S_{12}(q)$ for smaller values $\delta$ would
certainly be necessary to obtain a more precise estimate of this
exponent.

\section{Discussion}

We found that the left passage probability of zero-temperature domain
walls in the disordered SOS model in different geometries agrees well
(within the numerical error bars) with Schramm's formula for
$\kappa=4$. Since the fractal dimension of the domain walls is
$d_s=1.25\pm0.01$ this is inconsistent with $d_s=1+\kappa/8=1.5$ that
should hold if this ensemble of random curves would be described by
SLE. One condition for SLE to hold is that the measure of the ensemble
of random curves fulfills a domain Markov property. To check it
numerically is a computationally extremely challenging endeavour
(c.f.\ \cite{amoruso}) and we did not attempt it here.
The second condition for SLE is that the measure
is invariant under a conformal mapping of the domain within which the
random curves are defined. In our study we considered several
different domains shapes and we found that in all cases the left
passage probability is described by Schramm's formula adapted via a
conformal map to the specific geometry under consideration. Although
not being a sufficient criterion for conformal invariance to hold in
general this observation is at least surprising regarding the fact
that $\kappa$ and $d_s$ do not obey the SLE prediction.
To shed light on this issue further tests of conformal invariance
are worthwhile.

Interestingly the contour-loops in this system (i.e.\ the lines
connecting sites of equal height in the ground state) have a fractal
dimension close to $d_f=1.5$ \cite{zeng}, the same dimension as the
contour loops in the SOS model {\it without} disorder at finite
temperatures above the critical temperature \cite{cardy,sheffield}.
The latter {\it are} indeed described by SLE with $\kappa=4$
\cite{cardy,sheffield}. It remains to be checked, whether the
contour loops in the disordered SOS model are also also described by
SLE with $\kappa=4$.

The boundaries of droplets, i.e.\ connected clusters of a given
lateral size that have a minimal excitation energy, have the same
fractal dimension as domain walls ($d_s=1.25$), but their energy
$\Delta E_l$ saturates at a small, finite value for increasing lateral
size $l$.  In \cite{middleton-drop} it was shown that the mean droplet
energy $\Delta E_L$ decreases with system size $L$ like $\Delta
E_L^{-1}\sim\ln L$. This result is actually consistent with ours,
since in \cite{middleton-drop} the droplet size was only restricted by
an upper bound (the system size $L$) but not by a lower bound as in
our study. So the result for $\Delta E_L$ is the optimal excitation
among one of typically $\log L$ excitations of fixed (maximum and
minimum) size $l$. {\it If} the distribution of the energies of these
excitations on different length scales is independent of the length
scale, {\it then} $\Delta E_L$ is just the minimum of $\ln L$
independent, identically distributed random numbers, thus proportional
to $1/\ln L$. What we show in our study is essentially that this
assumption is indeed fulfilled, as shown in Fig.\ \ref{fig-exc-distr}.
This is a non-trivial result and does not immediately follow from a
vanishing stiffness exponent $\theta=0$. With a view towards their
entropic contribution one would like to know how the number of
independent (i.e.\ spatially disjoint) droplets scales with their
size. It should be possible to study these challenging questoin
with the methods that we presened.

Finally, we have shown that there is disorder chaos at $T=0$ in this
model. Our numerical data show a behavior of interlayer correlations
$S_{12}(q)$ compatible with Eq. (\ref{struct_tg}) with $\hat \sigma >
0$, thus rather similar to the one obtained close to $T_g$
\cite{hwa_fisher}. Here, the two ground states of the system, embedded in
two slightly different disorder realisations, thus display logarithmic
residual correlations. In addition, our data are consistent with a chaos
exponent $\alpha = 1$, in agreement with droplet scaling
argument. Following the studies of Ref. \cite{doussal-schehr, 
  pld_chaos, duemmer_chaos} it would certainly be interesting to describe
  analytically these behaviors by analysing in detail the model in
Eq. (\ref{frg_oneloopnonloc}) at $T=0$. Of course, especially in view of recent
analytical progress \cite{duemmer_chaos}, it would also be very
interesting to extend these studies of disorder chaos at finite
temperature, which we expect to play a role here in this marginal
glass phase.

\ack
We would like to thank P. Le Doussal for useful discussions.

\end{document}